\documentclass[twocolumn,prl,superscriptaddress,showpacs,amsmath]{revtex4}

\usepackage{graphicx}

\begin{document}

\title
{QND measurements and state preparation in quantum gases by light
detection}

\author{Igor B. Mekhov}
\email{Igor.Mekhov@uibk.ac.at} \affiliation{Institut f\"ur
Theoretische Physik, Universit\"at Innsbruck, Innsbruck, Austria}
\affiliation{St. Petersburg State University, Faculty of Physics,
St. Petersburg, Russia}
\author{Helmut Ritsch}
\affiliation{Institut f\"ur Theoretische Physik, Universit\"at
Innsbruck, Innsbruck, Austria}

\begin{abstract}
We consider light scattering from ultracold quantum gas in optical
lattices into a cavity. The measurement of photons leaking out the
cavity enables a quantum nondemolition (QND) access to various
atomic variables. The time resolved light detection projects the
motional state to various atom-number squeezed and macroscopic
superposition states that strongly depend on the geometry.
Modifications of the atomic and light properties at a single quantum
trajectory are demonstrated. The quantum structure of final states
can be revealed by further observations of the same sample.
\end{abstract}


\maketitle

Quantum gases in optical lattices are of fundamental interest, as
they provide an excellent testbed to study multipartite entanglement
and many-body states, useful in condensed matter and quantum
information~\cite{JakschBloch}. Usually, the role of light is
reduced to a classical auxiliary tool for creating intriguing atomic
states. In contrast, here we consider an ultimate quantum level,
where quantum natures of both matter and light play a key role. This
emerging level joining quantum optics [especially cavity quantum
electrodynamics (QED)] and quantum gases, only recently became
achievable and stimulated novel experimental
\cite{EsslReichCourtVulet} and theoretical studies
\cite{NatPhPrlPra,Meystre,EpjdNJPLewen}.

We show that the atom-light entanglement enriches physics and
enables the quantum nondemolition (QND) measurement and manipulation
of atomic states. Observing light allows to prepare different types
of atom number squeezed and macroscopic superposition states. Note
that the type of many-body states depends on the optical geometry.
The quantum structure of final states can be revealed by further
observations of the same sample, which is an advantage over
destructive schemes \cite{JakschBloch,Kasevich}.

As we consider off-resonant interaction, independent of a particle
level structure, our model might be also applied to other phenomena
in molecular physics~\cite{RempeMol}, where the molecule number
fluctuations are important, and solid-state systems as
semiconductors [Bose-Einstein condensates (BEC) of
exciton-polaritons]~\cite{LeSiDang} and
superconductors~\cite{Gambetta} (circuit cavity QED). Besides, the
squeezed and macroscopic superposition states find applications in
quantum interferometry and metrology \cite{GrangierPolzikCat}.

\begin{figure}
\scalebox{0.4}[0.4]{\includegraphics{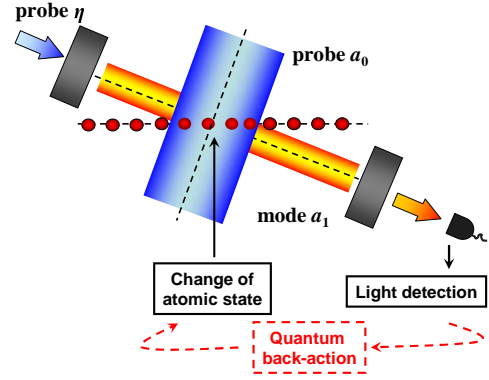}}
\caption{\label{fig1}(Color online) Setup. A lattice is illuminated
by the transverse probe $a_0$ and probe through a mirror $\eta$. The
photodetector measures photons leaking the cavity. Due to the
quantum back-action, the light measurement leads to the modification
of the atomic quantum state.}
\end{figure}

{\bf Model}. We consider (cf. Fig.~\ref{fig1}) $N$ ultracold atoms
in an optical lattice of $M$ sites formed by strong off-resonant
laser beams. A region of $K\le M$ sites is also illuminated by a
weak probe, which is scattered into a cavity. We will investigate,
how the measurement of photons leaking the cavity will affect the
atomic quantum state.

The theory is based on the generalized Bose-Hubbard model taking
into account the light quantization \cite{NatPhPrlPra,EpjdNJPLewen}.
In contrast to Ref.~\cite{EpjdNJPLewen}, we assume dynamics and
measurement of the cavity mode is faster than atomic
tunneling~\cite{NatPhPrlPra,Meystre}. Thus, neglecting the influence
of tunneling on light, we get the effective many-body Hamiltonian:
\begin{eqnarray}\label{1}
H=\hbar(\omega_1 + U_{11} \hat{D}_{11}) a^\dag_1 a_1+
\hbar U_{10}(\hat{D}^*_{10}a^*_0a_1 + \hat{D}_{10}a_0a^\dag_1)  \nonumber \\
-i\hbar(\eta^* a_1 - \eta a^\dag_1),
\end{eqnarray}
where $a_1$ is the cavity-mode annihilation operator and $a_0$ is
the c-number probe amplitude of the frequencies $\omega_{1,p}$ and
spatial mode functions $u_{1,0}({\bf r})$. $U_{lm}=g_lg_m/\Delta_a$
($l,m=0,1$), where $g_{1,0}$ are the atom-light coupling constants,
$\Delta_a=\omega_1 - \omega_a$ is the cavity-atom detuning, $\eta$
is the probe through a mirror at $\omega_p$. We assumed the
probe-cavity detuning $\Delta_{p}=\omega_{p}-\omega_1 \ll\Delta_a$.
The operators $\hat{D}_{lm}= \sum_{j=1}^K{u_l^*({\bf r}_j)u_m({\bf
r}_j)\hat{n}_j}$ sum contributions from all illuminated sites with
the atom-number operators $\hat{n}_j$ at the position ${\bf r}_j$.
Here, the rotating-wave approximation and adiabatic elimination of
the exited state were used.

The first term in Eq.~(\ref{1}) describes the atom-induced shift of
the cavity resonance. The second one reflects scattering
(diffraction) of the probe into a cavity. For a quantum gas the
frequency shift and probe-cavity coupling coefficient are operators,
which leads to different light scattering from various atomic
quantum states~\cite{NatPhPrlPra}.

The Hamiltonian (\ref{1}) describes QND measurements of variables
related to $\hat{D}_{lm}$ measuring the photon number $a^\dag_1a_1$
\cite{Brune}. Note, that one has a QND access to various many-body
variables, as $\hat{D}_{lm}$ strongly depend on the lattice and
light geometry via $u_{0,1}({\bf r})$. This is an advantage of the
lattice comparing to single- or double-well setups, where the photon
measurement back-action was considered \cite{BEC}. Moreover, such a
geometrical approach can be extended to other quantum arrays, e.g.,
ion strings \cite{ions}.

For example, $\hat{D}_{11}$ can reduce to the operator $\hat{N}_K$
of atom number at $K$ sites \cite{NatPhPrlPra}. If the probe and
cavity modes are coupled at a diffraction maximum (Bragg angle),
i.e., all atoms scatter light in phase, the probe-cavity coupling is
maximized, $\hat{D}_{10}=\hat{N}_K$. If they are coupled at a
diffraction minimum, i.e., neighboring atoms scatter out of phase,
$\hat{D}_{10}=\sum_{j=1}^K (-1)^{j+1}\hat{n}_j$ is the operator of
number difference between odd and even sites. Thus, the atom number
as well as number difference can be nondestructively measured. Note,
that those are just two of many examples of how a QND-variable, and
thus the projected state, can be chosen by the geometry.

{\bf Measurement back-action.} The expression for the initial
motional state of atoms reads
\begin{eqnarray}\label{2new}
|\Psi(0)\rangle =\sum_{q}c_q^0 |q_1,..,q_M\rangle,
\end{eqnarray}
which is a superposition of Fock states reflecting all possible
classical configurations $q=\{q_1,..,q_M\}$ of $N$ atoms at $M$
sites, where $q_j$ is the atom number at the site $j$. This
superposition displays the uncertainty principle, stating that even
a single atom can be delocalized in space.

While interacting, the light and atoms get entangled. Quantum
mechanics predicts that measurements of one subsystems (light)
provides conditional information about, or affects, another one
(gas). We will show, how the atomic uncertainty is affected by the
light detection.

We use the open system approach \cite{Carmichael} for counting
photons leaking the cavity of decay rate $\kappa$. When a photon is
detected, the jump operator is applied to the state:
$|\Psi_c(t)\rangle \rightarrow a_1|\Psi_c(t)\rangle$. Between the
counts, the system evolves with a non-Hermitian Hamiltonian
$H-i\hbar\kappa a^\dag_1a_1$. Such an evolution gives a quantum
trajectory for $|\Psi_c(t)\rangle$ conditioned on the detection of
photons at times $t_1,t_2,...$

It is known~\cite{DenisGardinerZoller} that, if a coherent probe
illuminates a classical atomic configuration in a cavity, the light
state is proportional to a coherent state $|\alpha_q(t)\rangle$ with
$\alpha_q(t)$ given by the classical Maxwell's equation. Thanks to
the approximation, where the tunneling does not affect light, we can
get a simple analytical solution of the coupled light-matter
dynamics. Each atomic Fock state in Eq.~(\ref{2new}) will be
correlated with a coherent light state with parameters given only by
the corresponding configuration $q$: $|\Psi_c(t)\rangle
=\sum_{q}c_q^0\exp[\Phi_q(t)]
|q_1,...,q_M\rangle|\alpha_q(t)\rangle/F(t)$, where $F(t)$ gives the
normalization. So, the problem to find $|\Psi_c(t)\rangle$ reduces
to finding $\alpha_q(t)$, $\Phi_q(t)$ for all classical
configurations forming the initial $|\Psi(0)\rangle$. Although a
solution is available for any $t$, we present it for $t>1/\kappa$,
when the steady state is achieved in all $\alpha_q(t)$, and assuming
the first photon was detected at $t_1>1/\kappa$.

Due to the steady state in all $\alpha_q(t)$, the solution is
independent of the detection times and after $m$ counts is
\begin{eqnarray}
|\Psi_c(m,t)\rangle =\frac{1}{F(t)}\sum_{q}\alpha_q^m e^{\Phi_q(t)}
c_q^0 |q_1,...,q_M\rangle|\alpha_q\rangle, \label{2}\\
\alpha_q=\frac{\eta-iU_{10} a_0D^q_{10}}{i(U_{11}
D^q_{11}-\Delta_p)+\kappa}, \label{3}\\
\Phi_q(t)=-|\alpha_q|^2\kappa t+(\eta\alpha^*_q-iU_{10}
a_0D^q_{10}\alpha^*_q-\text{c.c.})t/2, \label{4}
\end{eqnarray}
where $D^q_{lm}= \sum_{j=1}^K{u_l^*({\bf r}_j)u_m({\bf r}_j)q_j}$ is
a realization of $\hat{D}_{lm}$ at $\{q_1,..q_M\}$; $a_0$, $\eta$,
and $\alpha_q$ all oscillating in steady state at $\omega_p$ were
replaced by their constant amplitudes.

As we see, each light amplitude $\alpha_q(t)$, Eq.~(\ref{3}), is
given by a Lorentzian corresponding to classical optics.
Eq.~(\ref{2}) shows that the probability to find an atom
configuration $q$,
$p_q(m,t)=|\alpha_q|^{2m}\exp{(-2|\alpha_q|^{2}\kappa
t)}|c_q^0|^2/F^2$, changes in time due to the photodetection. This
demonstrates the back-action of the light measurement on the atomic
state.

In the following, we will show consequences of Eq.~(\ref{2}) for two
cases, where only one probe ($a_0$ or $\eta$) exists. For transverse
probing ($a_0\ne 0)$, we also neglect the mode shift, assuming
$U_{11} D^q_{11}\ll\kappa$ or $\Delta_p$. Thus, in both examples,
$\alpha_q$ (\ref{3}) depends on the configuration $q$ only via a
single statistical quantity now called $z$: $z=D^q_{11}$ for cavity
probing ($\eta \ne 0$), and $z=D^q_{10}$ for transverse probing.

From Eq.~(\ref{2}) we can determine the probability distribution of
finding a given $z$ after time $t$ as
\begin{eqnarray}\label{5}
p(z,m,t)=|\alpha_z|^{2m}e^{-2|\alpha_z|^2\kappa t}p_0(z)/F^2,
\end{eqnarray}
where the initial distribution $p_0(z)=\sum_{q'} |c_{q'}^0|^2$, such
that all configurations $q'$ have the same $z$; $F^2= \sum_z
|\alpha_z|^{2m}\exp{(-2|\alpha_z|^2\kappa t)}p_0(z)$ provides
normalization.

{\bf Transverse probing at diffraction maximum.} As was mentioned,
at the Bragg angle, $\hat{D}_{10}=\hat{N}_K$ is the operator of atom
number at $K$ sites. So, $z$ varies from $0$ to $N$ reflecting
possibilities to find any atom number at $K$ sites. The light
amplitudes (\ref{3}) $\alpha_z=Cz$ are proportional to $z$,
$C=iU_{10} a_0/(i\Delta_p-\kappa)$. The probability (\ref{5}) reads
\begin{eqnarray}\label{6}
p(z,m,t)=z^{2m}e^{-z^2\tau}p_0(z)/\tilde{F}^2
\end{eqnarray}
with a characteristic time $\tau=2|C|^2\kappa t$.

When time progresses, both $m$ and $\tau$ increase with a
probabilistic relation between them. The Quantum Monte Carlo method
\cite{Carmichael} establishes such a relation, thus giving a
trajectory. Note, that thanks to the simple analytical solution
(\ref{2}), it gets extremely simple. In each step, it consists in
the calculation of the photon number in the state given by
Eq.~(\ref{2}) and comparing it with a random number generated in
advance, thus, deciding whether the detection or no-count process
has happened.

\begin{figure}
\scalebox{1.0}[1.0]{\includegraphics{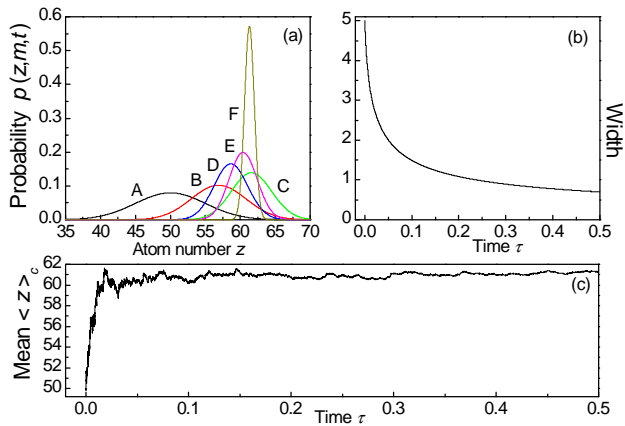}}
\caption{\label{fig2}(Color online) Photodetections at diffraction
maximum. (a) Shrinking atom number distribution at different times
$\tau=$ 0, 0.005, 0.018, 0.03, 0.05, 0.5 (A-F); (b) decreasing width
$\delta z$; (c) stabilizing mean atom number $\langle z\rangle_c$.
Initial state: SF, $N=100$ atoms, $K=M/2=50$ illuminated sites.}
\end{figure}

If the initial atom number $z$ at $K$ sites is uncertain, $p_0(z)$
is broad [for the superfluid (SF) it is nearly Gaussian
\cite{NatPhPrlPra}], and Eq.~(\ref{6}) shows that $p(z,m,t)$ is
strongly modified during the measurement. The function
$z^{2m}\exp{(-z^2\tau)}$ has its maximum at $z_1=\sqrt{m/\tau}$ and
full width at half maximum (FWHM) $\delta z \approx
\sqrt{2\ln2/\tau}$ (for $\delta z \ll z_1$). Thus, multiplying
$p_0(z)$ by this function will shrink the distribution $p(z,m,t)$ to
a narrow peak at $z_1$ with the width decreasing in time
(Fig.~\ref{fig2}).

This describes the projection of the atomic quantum state to a final
state with squeezed atom number at $K$ sites (a Fock states
$|z_1,N-z_1\rangle$ with $z_1$ atoms at $K$ sites and $N-z_1$ atoms
at $M-K$ sites). When $\delta z<1$, the final collapse is even
faster than $\sqrt{\tau}$, due to the discreteness of $p(z,m,t)$.
Measuring the photon number $m$ and time $t$, one can determine
$z_1$ of a quantum trajectory.

In contrast to recent results in spin squeezing \cite{PolzikNP},
which can be also obtained for thermal atoms
\cite{PolzikHotHolland}, in our work, quantum nature of ultracold
atoms is crucial, as we deal with the atom number fluctuations
appearing due to the delocalization of ultracold atoms in space.

After the distribution shrinks to a single $z_1$, the light
collapses to a single coherent state $|\alpha_{z_1}\rangle$, and the
atoms and light get disentangled with a factorized state
\begin{eqnarray}\label{7}
|\Psi_c\rangle=|z_1,N-z_1\rangle|\alpha_{z_1}\rangle.
\end{eqnarray}
So, light statistics evolves from super-Poissonian to Poissonian.
The conditioned (i.e., at a single trajectory) cavity photon number
$\langle a^\dag_1a_1\rangle_c(t)=|C|^2 \sum_{z=0}^Nz^2p(z,m,t)$ is
given by the second moment of $p(z,m,t)$. Its dynamics [very similar
to $\langle z\rangle_c$ in Fig.~\ref{fig2}(c)] has jumps, even
though all $\alpha_{z}(t)$ are continuous. In the no-count process,
$\langle a^\dag_1a_1\rangle_c$ decreases, while at one-count it
jumps upwards, which is a signature of super-Poissonian statistics.
Finally, it reduces to $\langle a^\dag_1a_1\rangle_c=|C|^2z_1^2$,
reflecting a direct correspondence between the final atom number and
cavity photon number, which is useful for experiments.

Even the final Fock state still contains the atom-atom entanglement,
as many components $|q_1,..,q_M\rangle$ can have the same $z_1$. For
example, the SF state can be represented as $|SF\rangle_{N,M}=\sum_z
\sqrt{B_z} |SF\rangle_{z,K}|SF\rangle_{N-z,M-K}$ ($B_z$ are binomial
coefficients). After the measurement, it ends up in
$|SF\rangle_{z_1,K}|SF\rangle_{N-z_1,M-K}$, i.e., the product of two
uncorrelated superfluids.

Our measurement scheme determines (by squeezing) the atom number at
a particular lattice region and projects the initial atomic state to
some subspace. However, the atom number at different regions keeps
quantum uncertainty. So, the quantum structure of the final state
can be revealed in a further optical or matter-wave experiment.
Thanks to the lattice geometry, one can change the illuminated
region, and further study measurement-induced collapse of the state
in the remaining subspace.

Even in matter-wave experiments \cite{JakschBloch}, the product of
SFs will look different from the initial SF: the atoms from
different regions will not interfere in average. Note, that we did
not specify how $K$ sites were selected. One can illuminate a
continuous region. However, one can illuminate each second site by
choosing the probe wavelength twice as lattice period and get number
squeezing at odd and even sites. In this way, one gets a
measurement-prepared product of two SFs ``loaded'' at sites one by
one (e.g. atoms at odd sites belong to one SF, while at even sites
to another). While the initial SF shows the long-range coherence
$\langle b^\dag_i b_j\rangle$ with the lattice period, the prepared
state will demonstrate the doubled period in $\langle b^\dag_i
b_j\rangle$ ($b_j$ is the atom annihilation operator).

{\bf Transverse probing at diffraction minimum.} In contrast to
classical atoms, quantum gases scatter light even in diffraction
minima \cite{NatPhPrlPra}. Here $z=D^q_{10}=\sum_{j=1}^M
(-1)^{j+1}q_j$ is the atom number difference between odd and even
sites, varying from $-N$ to $N$ with a step 2 (we assumed $K=M$).
Eq.~(\ref{6}) keeps its form with a new meaning of $z$ and $p_0(z)$
[for SF, new $p_0(z)$ is nearly a Gaussian centered at $z=0$ and the
width $\sqrt{N}$ \cite{NatPhPrlPra}].

\begin{figure}
\scalebox{1.0}[1.0]{\includegraphics{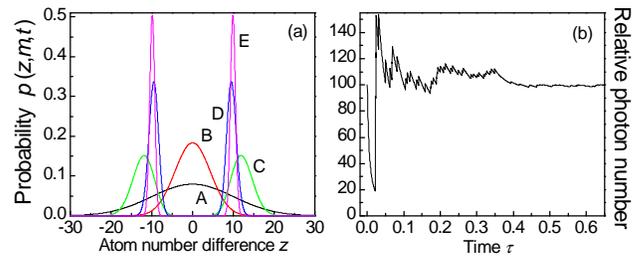}}
\caption{\label{fig3}(Color online) Photodetections at diffraction
minimum. (a) Shrinking  distribution of the atom-number difference
for various times $\tau=$ 0, 0.02, 0.03, 0.17, 0.65 (A-E). The
doublet corresponds to macroscopic superposition state. (b) Relative
photon number $\langle a^\dag_1 a_1\rangle_c/|C|^2$ with quantum
jumps. Initial state: SF, $N=100$ atoms, $K=M=100$ sites.}
\end{figure}

The striking difference from the diffraction maximum is that our
measurement (\ref{6}) is not sensitive to the sign of $z$, while the
amplitudes $\alpha_z=Cz$ are. So, the final state is a macroscopic
superpositions of two Fock states with $z_{1,2}=\pm \sqrt{m/\tau}$
and light amplitudes: $\alpha_{z_2}=-\alpha_{z_1}$,
\begin{eqnarray}\label{8}
|\Psi_c\rangle=(|z_1\rangle|\alpha_{z_1}\rangle+
(-1)^m|-z_1\rangle|-\alpha_{z_1}\rangle)/\sqrt{2}.
\end{eqnarray}
Figure \ref{fig3} shows the collapse to a doublet probability $p(\pm
z_{1},m,t)$ and the photon-number trajectory, where upward jumps and
no-count decreases can be seen.

In contrast to a maximum, even in the final state, the light and
matter are not disentangled. Moreover, to keep the purity of the
state, one should know precisely the number of detected photons,
because of the sign flip in Eq.~(\ref{8}). This reflects the
fragility of macroscopic superposition states with respect to the
decoherence.

As a result, the measurement-based state preparation at the
diffraction maximum (\ref{7}) is much more robust.

For probing through a mirror ($\eta\ne 0$, $a_0=0$), in contrast to
the transverse probing, the probability distribution can collapse
both to a singlet and doublet. Importantly, here the superposition
state can be more robust than (\ref{8}). This is due to a smaller
phase jump between two Fock states, which can be obtained from
Eqs.~(\ref{2})-(\ref{4}).

Deviations from our ideal setup can slightly modify the results,
i.e., the mode profiles will lead, instead of atom number, to more
general variables given by $\hat{D}_{lm}$.

In summary, we showed that using the light-matter entanglement in
ultracold gases enables QND measurements of different atomic
variables and creation of specific atomic states. The state type is
determined by the optical geometry. Our model can be generalized to
other quantum arrays. Cavity QED with quantum gases can operate with
atom numbers from millions to one \cite{Rempe}. Thanks to recent
experimental breakthroughs \cite{EsslReichCourtVulet}, preparing
various kinds of atom number squeezing is already doable, and
creation of the superposition states with, at least, small particle
number \cite{GrangierPolzikCat} may become practical.

We thank D. Ivanov, E. Polzik, and A. Vukics for stimulating
discussions. Support by FWF (P17709, S1512).

\end{document}